\documentclass[conference]{IEEEtran}
\IEEEoverridecommandlockouts
\usepackage{cite}
\usepackage{amsmath,amssymb,amsfonts}
\usepackage{algorithm}
\usepackage{algorithmic}
\usepackage{booktabs}
\usepackage{graphicx}
\usepackage{subfigure}
\usepackage{textcomp}
\usepackage{xcolor}
\usepackage{graphicx}
\usepackage{tablefootnote}
\usepackage{multirow}
\usepackage[colorlinks,
            linkcolor=blue,
            anchorcolor=blue,
            citecolor=blue]{hyperref}
\def\BibTeX{{\rm B\kern-.05em{\sc i\kern-.025em b}\kern-.08em
    T\kern-.1667em\lower.7ex\hbox{E}\kern-.125emX}}
\begin{document}

\title{DQR-TTS: Semi-supervised Text-to-speech Synthesis with Dynamic Quantized Representation\\
}

\author{\IEEEauthorblockN{Jianzong Wang$^{1}$, Pengcheng Li$^{1,2}$, Xulong Zhang$^{1\ast}$\thanks{$^\ast$Corresponding author: Xulong Zhang (zhangxulong@ieee.org).}, Ning Cheng$^{1}$, Jing Xiao$^{1}$}
\IEEEauthorblockA{\textit{$^{1}$Ping An Technology (Shenzhen) Co., Ltd.}\\\textit{$^{2}$University of Science and Technology of China}}
}

\maketitle

\begin{abstract}
Most existing neural-based text-to-speech methods rely on extensive datasets and face challenges under low-resource condition. In this paper, we introduce a novel semi-supervised text-to-speech synthesis model that learns from both paired and unpaired data to address this challenge. The key component of the proposed model is a dynamic quantized representation module, which is integrated into a sequential autoencoder. When given paired data, the module incorporates a trainable codebook that learns quantized representations under the supervision of the paired data. However, due to the limited paired data in low-resource scenario, these paired data are difficult to cover all phonemes. Then unpaired data is fed to expand the dynamic codebook by adding quantized representation vectors that are sufficiently distant from the existing ones during training. Experiments show that with less than 120 minutes of paired data, the proposed method outperforms existing methods in both subjective and objective metrics.
\end{abstract}


\section{Introduction}
The objective of text-to-speech (TTS) is to produce coherent and lifelike speech from provided textual content. It has been broad used in various areas~\cite{xu2021survey,sung2022meta}, including voice assistants, telephone services, vedio games, \textit{etc}. 

Cascaded TTS systems commonly employ a pipeline that comprises an acoustic model and a vocoder, with mel spectrograms or other linguistic features serving as the intermediate representations~\cite{wang2017tacotron,jonathan2018natural}. Recently, many neural network-based end-to-end TTS models like Fastspeech 2s~\cite{ren2020fastspeech}, EATS~\cite{donahue2020end}, VITS~\cite{kim2021conditional}, \textit{etc.} have emerged, which not only enhance the accuracy and clarity of synthesized speech but also make significant strides in achieving a more natural and human-like sound quality. However, the success of most neural network-based TTS models always relies on the extensive and high-fidelity training data. In terms of data quality, it is crucial that the audio content covers an adequate range of phonemes, and the distribution of these phonemes should be carefully balanced. On the other side, training a high-performing TTS model necessitate a significant volume of paired data (\textit{i.e.} record along with its corresponding transcript), which can be expensive and time-consuming to label manually\cite{zhang2022TDASS}. Exploring the utilization of unpaired data (\textit{i.e.} only speech data) for training or enhancing TTS models is worthy of research.

Weakly supervised or unsupervised learning has been applicated for TTS in some methods \cite{alexander2019towards,nachmani2019un,zhang2020un,chorowski2019unsupervised,ni2022unsupervised}. An almost unsupervised learning approach for TTS by incorporating dual transformation and bidirectional sequence modeling was introduced in \cite{ren2019almost}. The main concept involves leveraging an automatic speech recognition (ASR) model to generate pseudo text annotations, thereby converting unpaired data into paired data. Chung \textit{et al.}\cite{chung2019semi} encapsulate each word in the input text with word vectors and incorporate them into the Tacotron \cite{wang2017tacotron} encoder. Subsequently, they employ an unpaired speech corpus to pre-train the Tacotron decoder within the acoustic domain, followed by fine-tuning the model using the accessible paired data.

Nevertheless, these TTS methods suffer the following shortcomings: (1) Some methods rely heavily on the pre-trained ASR, the quality of pseudo-labels has a significant impact on training. (2) These methods face challenges in explicitly representing acoustic characteristics, as they may not fully leverage the potential of unpaired data. 

In response to these constraints, we introduce a novel semi-supervised TTS model with \textbf{D}ynamic \textbf{Q}uantized \textbf{R}epresentation called \textbf{DQR-TTS}, which learns from paired and unpaired data. When paired data is provided, the dynamic codebook learns the quantized representations in a supervised way. After that, unpaired data are leveraged to expand the dynamic codebook through a designed learning strategy. This paper's contributions can be outlined as follows:
\begin{itemize}
    \item An autoencoder with a dynamic codebook is proposed, which can learn from paired data and expand the codebook from low-quality data based on a designed learning strategy to cover a wider range of phonemes.
    \item The semi-supervised TTS model is capable of addressing the challenges of low-resource scenarios, and not rely on accurate pseudo labels. Experiments show that DQR-TTS achieves desirable performance with limited paired data.
\end{itemize}

\begin{figure*}[t]
\centering
\centerline{\includegraphics[width=0.88\textwidth]{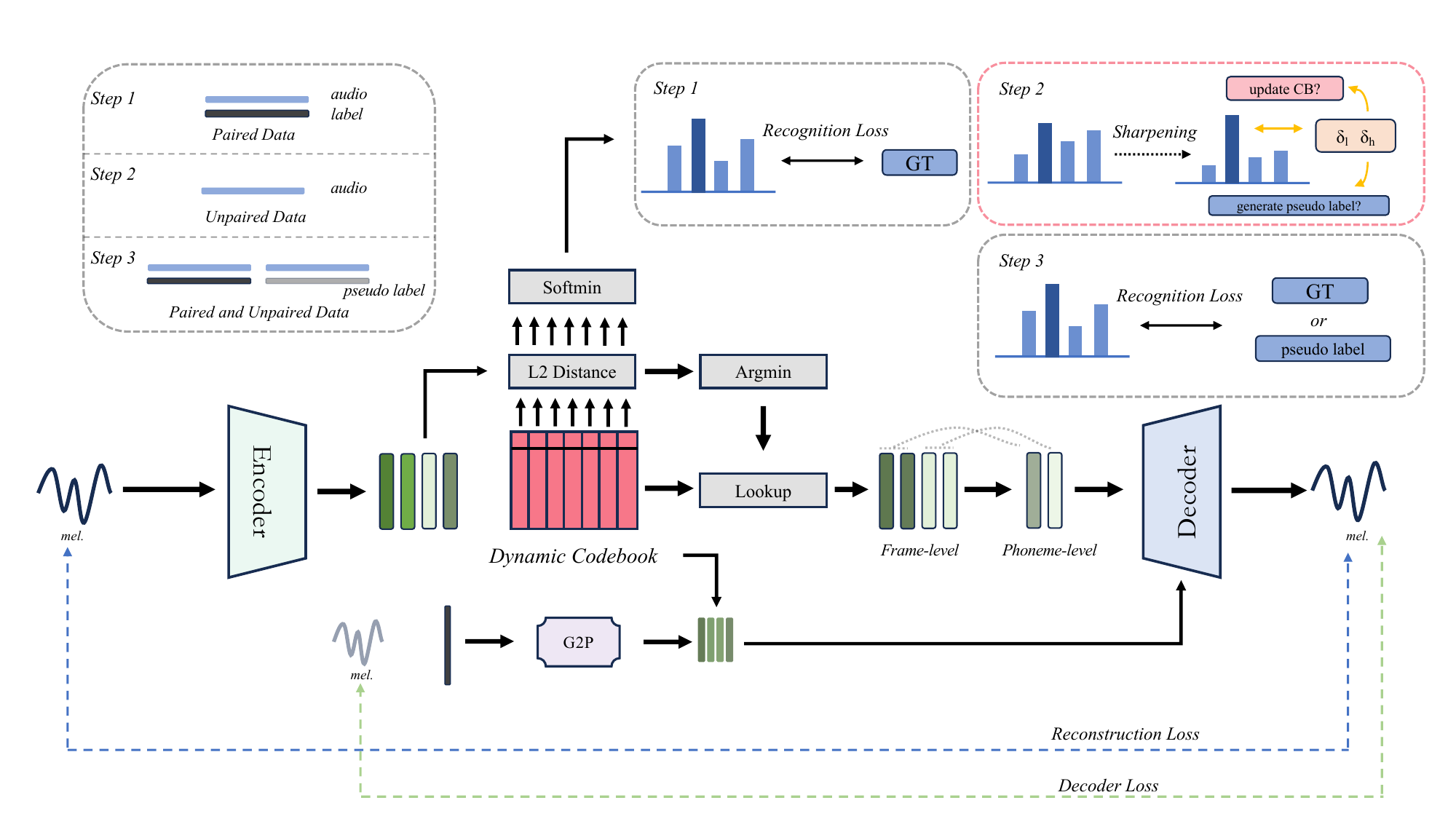}}
\caption{Pipeline of DQR-TTS. The training process consists of three steps: (1) The paired data is fed into the network to compare the distance between each continuous vector and codeword, and then replace each vector with the nearest codeword respectively. Then Mapping the codewords to phonemes according to the labels. (2) Generate pseudo labels via pre-trained ASR and execute the dynamic codebook updating strategy with unpaired data. (3) Mapping the unpaired codewords to phonemes according to the pseudo labels. Then the paired data and unpaired data with pseudo labels are jointly utilized to train the model.}
\label{network_structure}
\end{figure*}

\section{Related Work}
\subsection{Speech Representation Quantization}
Numerous prior studies have emphasized the acquisition of representations with continuous features \cite{vincent2010stacked,denton2016semi,chen2016infogan}. Nevertheless, discrete representations offer advantages more suitable for planning or complex reasoning\cite{mnih2014neural,oord2017neural}. Oord \textit{et al.} \cite{oord2017neural} introduce a method combines vector quantization (VQ) and variational autoencoders (VAE) \cite{kingma2014auto} to train the autoencoders with discrete hidden variables.

In the field of speech, learning discrete representation has been used for different tasks. Several works leverage VQ for voice conversion \cite{wu/2020/vqvc,wu/2020/vqvcplus,tang2022AVQVC,Tang2023VQ-CL}, codebook is leveraged to extract content representation from source speech. Liu \textit{et al.}\cite{liu2020towards} design an autoencoder to efficiently learn from speech data then perform text-to-speech synthesis. This model is capable of generating sequence of representations that closely resemble the phoneme sequence of speech utterances. Additionally, they have extended this method to multi-speaker TTS via a speaker representation table \cite{tu2020semi}. Another work \cite{liu2023dse} demonstrates that VQ acoustic feature is more suitable for cross-lingual text-to-speech (CTTS), and proposes a framework based on VQ, which consists of text-to-vec and vec-to-wav two stages.

\subsection{Semi-supervised Text-to-speech}
For some low-resource scenarios, such as TTS in less common languages, the cost of creating large paired datasets is prohibitively high, so researchers have turned to applying semi-supervised learning to TTS. The semi-supervised approach offers a more flexible approach to enhance low-resource TTS systems \cite{shechtman2019sequence,khar2023speak,zhang2022Semi}, which holds business value and contributes to social welfare \cite{wang2018segmental}. Inoue \textit{et al.} \cite{inoue2020semi} leverage an ASR trained with limited paired data to generate low-accuracy pseudo labels for unpaired data. Then TTS is pre-trained on the generated corpus, next to be fine-tuned. Guo \textit{et al.} \cite{guo2023qstts} propose a semi-supervised TTS model, which combines multi-stage and multi-codebook (MSMC) framework and self-supervised representation learning to improve its synthesis quality.

Our proposed method is based on dynamic quantized representation, which includes a dynamic codebook. It incorporates a designed dynamic representation learning strategy in a semi-supervised manner, enabling it to flexibly address the low-resource challenge.

\section{Methodology}
\subsection{Sequential AutoEncoder}
The pipeline of DQR-TTS is shown in Fig. \ref{network_structure}. Given an input speech $X$, we first divide it into $T$ frames, so the input can be present as $X=\{x_1, x_2, ..., x_T\}$. The encoder $Enc({\cdot})$ with parameter $\theta$ extracts frame-level hidden representation vectors $z_t^f$ from $X$ as:
\begin{equation}
    Enc_{\theta}(X)=\{z^{f}_{1}, z^{f}_{2}, ..., z^{f}_{T}\}
\end{equation}
where $z_{t}^{f} \in \mathbb{R}^{D_f}$, $D$ denotes the dimension of hidden presentation vector and $f$ denotes frame-level based.

The proposed dynamic phonemic representation module (DPRM) quantizes the frame-level hidden representation, and then clusters the frame-level vectors that represent the same phoneme into a phoneme-level vector $z^p_S$. This process will be illustrated in the next section. 

The decoder $Dec({\cdot})$ with parameter $\sigma$ reconstructs the speech from the phoneme-level vectors to ensure that these vectors adequately represent the original input speech. The reconstructed speech $X'$ is as follows:
\begin{equation}
\label{eq:recon}
   X'=Dec_{\sigma}(\{z^p_1, z^p_1, ..., z^p_S\})
\end{equation}

The reconstruction loss describes the distinctions between original speech $X$ and reconstruct speech $X'$ on the frame-level:
\begin{equation}
   \mathcal{L}_{recon}=||X - X'||_{2}
\end{equation}

\subsection{Quantized Representation}
\subsubsection{Frame-synchronized Quantization}
As the encoder generates continuous frame-level hidden representation vectors which are difficult to interpret, as multiple adjacent vectors may represent the same phoneme. To address this issue, we cluster these continuous vectors into groups and assign each vector a specific type for identification. Thus, we introduce a codebook $B=\{b_1, b_2, b_3, ..., b_n\}$ to represent the certain vectors with the chosen codeword $b_i \in B$. Then the output of encoder $Z^f=\{z^{f}_{1}, z^{f}_{2}, ..., z^{f}_{T}\}$ is replaced by the codeword $b_n$ in the codebook according to $L2$ distance:
\begin{equation}
\label{equ:codeword}
    z^{f}_{t}=\mathop{\arg\min}_{b_i}||z^{f}_{t}-b_i||_2
\end{equation}

Due to the operation in Eq. \ref{equ:codeword} being non-differentiable, we employ the technique outlined in \cite{yoshua2013estimating} to estimate the gradient of this process:
\begin{equation}
    {\bar{z}}^{f}_{t} = {z}^{f}_{t}+b_n-sg({z}^{f}_{t})
\end{equation}
where $sg(\cdot)$ is the stop-gradient operation which considers its input as unchanging during back-propagation and ${\bar{z}}^{f}_{t}$ denotes the quantized frame-level vector.

\subsubsection{Phoneme-synchronized Segmentation}
Temporal segmentation for continuous signals poses a challenge, but it becomes more operable with the introduction of vector quantization (\textit{i.e.} frame-synchronized quantization). The \textit{frame-synchronized quantization} operation replaces all frame-level vectors generated by the encoder with codewords from the codebook, this limits each $\bar{z}^f_t$ to a finite number of possibilities, which is related to the size of the codebook. Then we conduct \textit{phoneme-level combination} to merge the same and adjacent codewords. This operation converts the representation vectors from frame-level to phoneme-level, which ensures each vector represents a phoneme:
\begin{equation}
    Ph_{comb}(\{\bar{z}^{f}_{1}, \bar{z}^{f}_{2}, ..., \bar{z}^{f}_{T}\})=\{z^p_1, z^p_2, ..., z^p_S\}
\end{equation}
where $Ph_{comb}(\cdot)$ denotes the \textit{phoneme-level combination} operation, the quantity of vectors is compressed from $T$ to $S$. We compute the mean of each cluster of frame-level vectors to make the model training stable. Therefore each representation vector $z^p_i$ corresponds to a phonetic unit, as each entry in the codebook is associated with a phoneme, so we obtain phoneme-synchronized representations from frame-level audio sequence.

\subsection{Semi-supervised Dynamic Representation Learning}
The training data of our model contents only a limited quantity of paired data $(X_{pair}, Y_{pair})$ and a substantial quantity of unpaired data $X_{unpair}$, where $X_{pair}$ and $X_{unpair}$ denote audio sequences while $Y_{pair}$ is the corresponding label that records the phoneme sequence of $X_{pair}$. The representations of the phoneme sequence $Y_{pair}$ or pseudo phoneme sequence label $Y_{pseudo}$ can be built according to the codebook. Our proposed method introduces a \textit{dynamic learning strategy} for the dynamically updating codebdatook, allowing it to capture phoneme representations from both paired and unpaired data. 

Firstly, we put all paired data into the network to perform reconstruction. The probability of vector $z^{f}_{t}$ being mapped to a codeword $b_n$, which will be mapped to a phoneme, is formally characterized as:
\begin{equation} \label{prob}
    P(b_{n}|z^{f}_{t})=\frac{exp(-||z^{f}_{t}-b_n||_{2})}{\sum_{k\in N}exp(-||z^{f}_{t}-b_k||_{2})}
\end{equation}
then the probability for a frame-level phoneme sequence $\hat{Y}=(b_1,b_2,...,b_T)$ is approximated by:
\begin{equation}
    P(\hat{Y}|Enc_{\theta}(X))=\prod_{n=1}^{T}P(b_{n}|z^{f}_{n})
\end{equation}
The connectionist temporal classification (CTC) \cite{alex2006conn} is leveraged to address the mismatch in length between the $\hat{Y}$ (length $T$) and $Y_{pair}$ (length $S$). Then, each vector representation in the codebook can be mapped to a phoneme according to $\hat{Y}$. The recognition loss can be written as: 
\begin{equation}
\label{eq:recog}
   \mathcal{L}_{recog}=-\log{P(Y_{pair}|Enc_{\theta}(X))}
\end{equation}

Note that only performing reconstruction and recognition cannot get the relations between codewords and phonemes. As the training goes on, we conduct mapping codewords in the codebook to phonemes with a grapheme-to-phoneme converter (G2P) converting text labels to phoneme sequences. 

Secondly, we leverage an ASR model to generate pseudo labels for unpaired data. Then put unpaired data into the model and allow the codebook to enlarge like the previous step but follows a strategy, for the pseudo labels are not accurate. Since the probability of unpaired data may have a low entropy, we sharpen it with temperature $\tau$:

\begin{equation}
     P_{un}(b_{n}|z^{f}_{t})=\frac{exp(-||z^{f}_{t}-b_n||_{2}/\tau)}{\sum_{k\in N}exp(-||z^{f}_{t}-b_k||_{2}/\tau)}
\end{equation}

We introduce the algorithm as shown in Alg. \ref{alg} to enhance the dynamic codebook, expanding its range of phonemes and improving its representation capability. 


\begin{algorithm}[!h]
    \caption{Dynamic Codebook Update}
    \label{alg}
    \renewcommand{\algorithmicrequire}{\textbf{Input:}}
    \begin{algorithmic}[1]
        \REQUIRE audio seq. $X_{unpair}$, thresholds $\delta_l$ and $\delta_h$
        \STATE $Z=Enc_{\theta}(X_{unpair})$
        \FOR{each $z^{f}_{t} \in Z$}
            \FOR{each $b_n \in B$}
            \STATE compute $P_{un}(b_{n}|z^{f}_{t})$
            \ENDFOR
            \STATE get the max ${P_{un}}(b_n|z^{f}_{t})$ as $\hat{P_{un}}(b_n|z^{f}_{T})$
            \IF{$\hat{P_{un}}(b_n|z^{f}_{T}) $\textless$ \delta_l$}
                \STATE add $z^f_T$ to the codebook
            \ELSIF{$\hat{P_{un}}(b_n|z^{f}_{T}) $\textgreater$ \delta_h$}
                \STATE refine the pseudo label
            \ELSE
                \STATE drop out and \textbf{continue}
            \ENDIF
        \ENDFOR
    \end{algorithmic}
\end{algorithm}

\begin{itemize}
\item $\hat{P_{un}}(b_{n}|v^{f}_{T})<\delta_l$: this vector representation has  never been presented in the paired data, and could be a new phoneme representation in the unpaired data.
\item $\hat{P_{un}}(b_{n}|v^{f}_{T})>\delta_h$: the vector representation certainly enough to be representation as a phoneme in the codebook, which means this phoneme has already been represented in the paired data.
\item $\delta_l<\hat{P_{un}}(n|v^{f}_{T})<\delta_h$: difficult to determine whether it can become a new codeword or already exists in the codebook.
\end{itemize}

However, there may exists codewords from unpaired data which not appeared in paired data. So we also use the pseudo labels to map additional codewords to phonemes. We give priority to using accurate labels from paired data for mapping as in step 1. This reducing the reliance on the accuracy ASR model.

Finally, we leverage the paired data and unpaired data with pseudo labels to train the model jointly with the decoder loss included. The decoder in DQR-TTS is founded upon Tacotron 2 \cite{jonathan2018natural}, so another loss component is required during training:
\begin{equation}
\label{eq:dec}
    \mathcal{L}_{dec}=||Dec_{\sigma}(Ph) - X||_{2}
\end{equation}
where $Ph$ denotes the phoneme sequence retrieve from label or pseudo label according to the codebook. The overall joint training loss function of the proposed method in Eq. \ref{eq:loss} is a combination of reconstruction loss in Eq. \ref{eq:recon}, recognition loss in Eq. \ref{eq:recog}, as well as the decoder loss in Eq. \ref{eq:dec}. 

\begin{equation}
\label{eq:loss}
   \mathcal{L}_{total} = \mathcal{L}_{recon} + \alpha_1 \cdot \mathcal{L}_{recog} + \alpha_2 \cdot \mathcal{L}_{dec}
\end{equation}

\section{Experiments and Results}
\subsection{Experiment Setup}
To evaluate our proposed DQR-TTS, we conduct experiments on LJSpeech \cite{ljspeech17} corpus. The single-speaker dataset contains about 24 h audio. The dataset provides the paired data required for the experiment. As for the unpaired data used in the semi-supervised training, we select a portion of it from the dataset and ignore its transcriptions, treating it as unpaired data. The 50 ms window and the 12.5 ms hop size perform spectrogram extraction. For the linguistic units, we leverage a G2P converter\footnote[1]{\url{https://github.com/Kyubyong/g2p}} to complete the generation of phoneme sequence. The encoder in DQR-TTS simply consists of convolution blocks and LSTMs \cite{sepp1997long}, while the decoder is based on Tacotron 2 \cite{jonathan2018natural}. We use a WaveNet\cite{oord2016wavenet}-based vocoder which is pre-trained to convert spectrogram into time-domain waveform. 

In the experiments, DQR-TTS is trained with an Adam optimizer \cite{kingma2015adam} (with $\beta_1=0.9$, $\beta_2=0.999$, $\epsilon=10^{-6}$, \textit{lr}=$10^{-3}$). A batch size of 64 is employed during the model training. In our experiments, we set $\alpha_1$ and $\alpha_2$ in Eq. \ref{eq:loss} to $0.5$ and $1$ respectively.

As for evaluation metrics, we use both objective and subjective assessment methods to assess the fidelity of audio. We use Mean Opinion Score (MOS) as the subjective metric, 16 participants are invited to rate the synthesized speech, score ranges from 1 to 5, where a higher score signifies superior speech quality and naturalness. Mel-Cepstral Distortion (MCD) and Phoneme Error Rate (PER) are leveraged as objective metrics. MCD computes the spatial separation among the synthesized speech and the GT speech. PER calculates the error rate of phonemes in the synthesized speech according to the target label.

\subsection{Comparing with Other Methods}
We compare our proposed method with Tacotron 2 \cite{jonathan2018natural}, Speech Chain \cite{tjandra2017listening}, SeqRQ-AE (baseline method) \cite{liu2020towards}, and UASR-TTS \cite{ni2022unsupervised}. 
We randomly sample 20 sentences from the test set for performing the MOS and MCD test. The baseline method and the proposed DQR-TTS are semi-supervised, we train them on 120 min paired data and 600 min unpaired data, while UASR-TTS is unsupervised, we train it with the same amount of data. Tacatron 2 is a fully-supervised TTS model and is widely used, we train it with 12 h paired data. As for the Speech Chain, which is a dual learning framework encompassing both ASR and TTS without shared representations, we carry out the training of ASR and TTS modules within this framework using paired data and pseudo-paired data. The results of models trained with about 12 hours
data (paired, unpaired, or mix) are presented in Table \ref{tab:1}.
\begin{table}[htb]
\centering
\caption{Comparison of Different Methods}
\label{tab:1}
\begin{tabular}{cccc}
\toprule
{Method}&{Training Data\tablefootnote{The two components represent paired and unpaired data respectively.}}&{MOS$\uparrow$}&{MCD$\downarrow$}\\
\midrule
{Ground truth}& -  & 4.81 $\pm$ 0.07 & - \\
\midrule
{Tacotron 2}& 720 + 0 min & 3.54 $\pm$ 0.12 & 4.17 $\pm$ 0.03\\
{Speech Chain}& 120 + 600 min & 1.92 $\pm$ 0.57 & 8.61 $\pm$ 0.16\\
{SeqRQ-AE}& 120 + 600 min & 2.85 $\pm$ 0.22 & 4.83 $\pm$ 0.06\\
{UASR-TTS}& 0 + 720 min & 3.04 $\pm$ 0.27 & 5.32 $\pm$ 0.06\\
\midrule
{\bfseries DQR-TTS}& 120 + 600 min & 3.12 $\pm$ 0.32 & 4.54 $\pm$ 0.04\\
\bottomrule
\end{tabular}
\end{table}

We also conduct experiments under a simulated low-resource condition, which means that the data (both paired and unpaired) for training is limited. The results are showcased in Table \ref{tab:2}. Reducing the quantity of training data significantly impacts the speech synthesis performance of TTS models, especially for fully-supervised model. Our proposed model outperforms other supervised or unsupervised models in the low-resource scenario.
\begin{table}[htb]
\centering
\caption{Comparison of Different Methods in Low-Resource Scenario}
\label{tab:2}
\begin{tabular}{cccc}
\toprule
{Method}&{Training Data}&{MOS$\uparrow$}&{MCD$\downarrow$}\\
\midrule
{Ground truth}& -  & 4.81 $\pm$ 0.07 & - \\
\midrule
{Tacotron 2}& 120 + 0 min & 2.35 $\pm$ 0.39 & 6.42 $\pm$ 0.15\\
{Speech Chain}& 120 + 0 min & 1.62 $\pm$ 0.13 & 8.23 $\pm$ 0.43\\
{SeqRQ-AE}& 60 + 60 min & 2.58 $\pm$ 0.20 & 7.18 $\pm$ 0.46\\
{UASR-TTS}& 0 + 120 min & 2.64 $\pm$ 0.31 & 6.73 $\pm$ 0.11\\
\midrule
{\bfseries DQR-TTS}& 60 + 60 min & 2.83 $\pm$ 0.27 & 6.29 $\pm$ 0.18\\
\bottomrule
\end{tabular}
\end{table}

\subsection{Training with Different P/U Data Ratio}
To verify the capacity of DQR-TTS and the baseline method in different data volume scenarios, we conduct experiments on paired and unpaired data in different lengths. For our experiment, We limit the amount of unpaired data in the training set to 300 min and train it in conjunction with varying quantities of paired data. Training is performed under the conditions where the ratios of paired data to unpaired data are $1:10$, $1:5$, $1:2.5$, and $1:1$, respectively.

\begin{figure}[htb]
\centering 
\subfigure[Proposed Method] { 
\includegraphics[width=0.85\columnwidth]{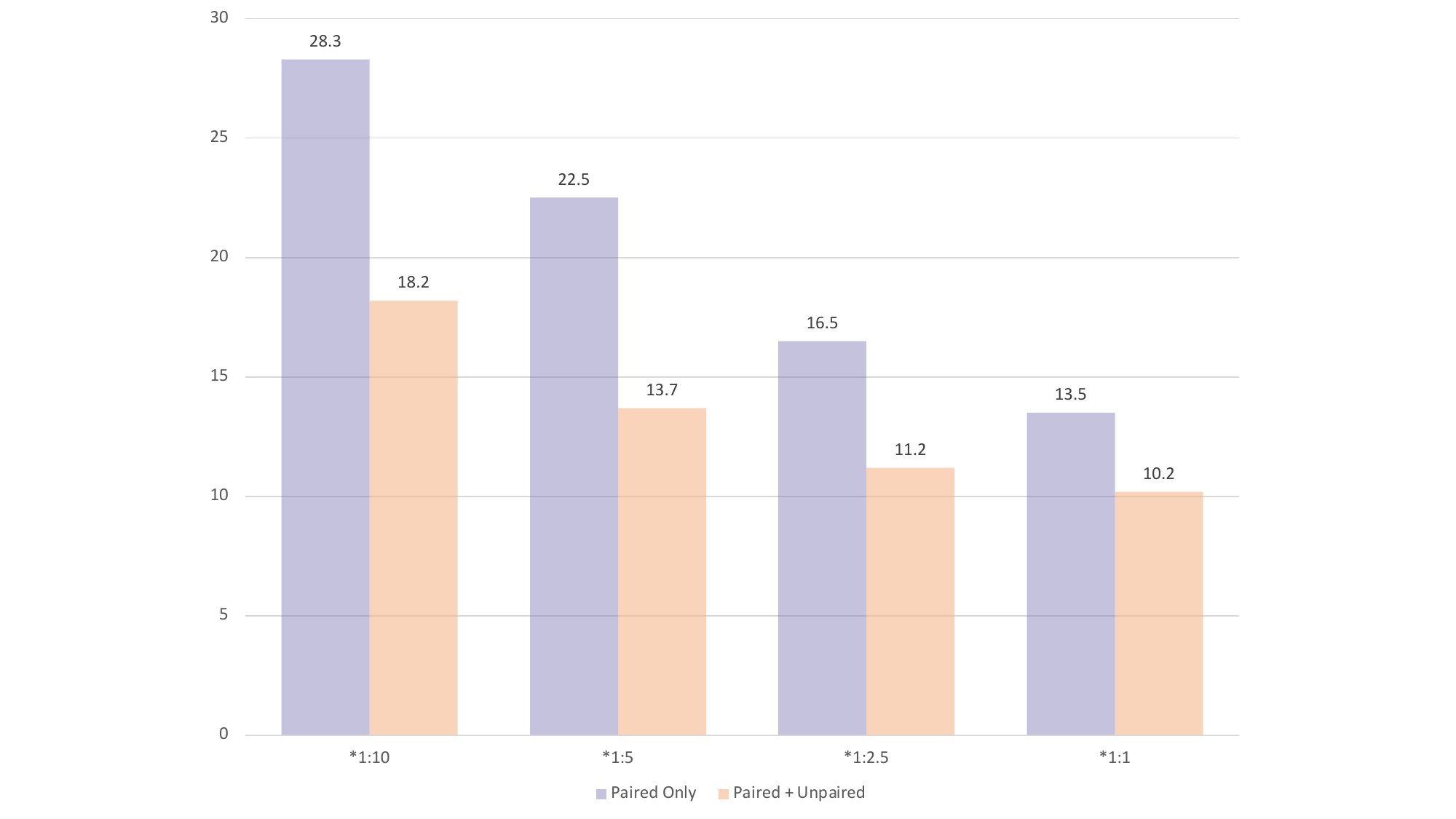} 
} 
\subfigure[Baseline Method] { 
\includegraphics[width=0.95\columnwidth]{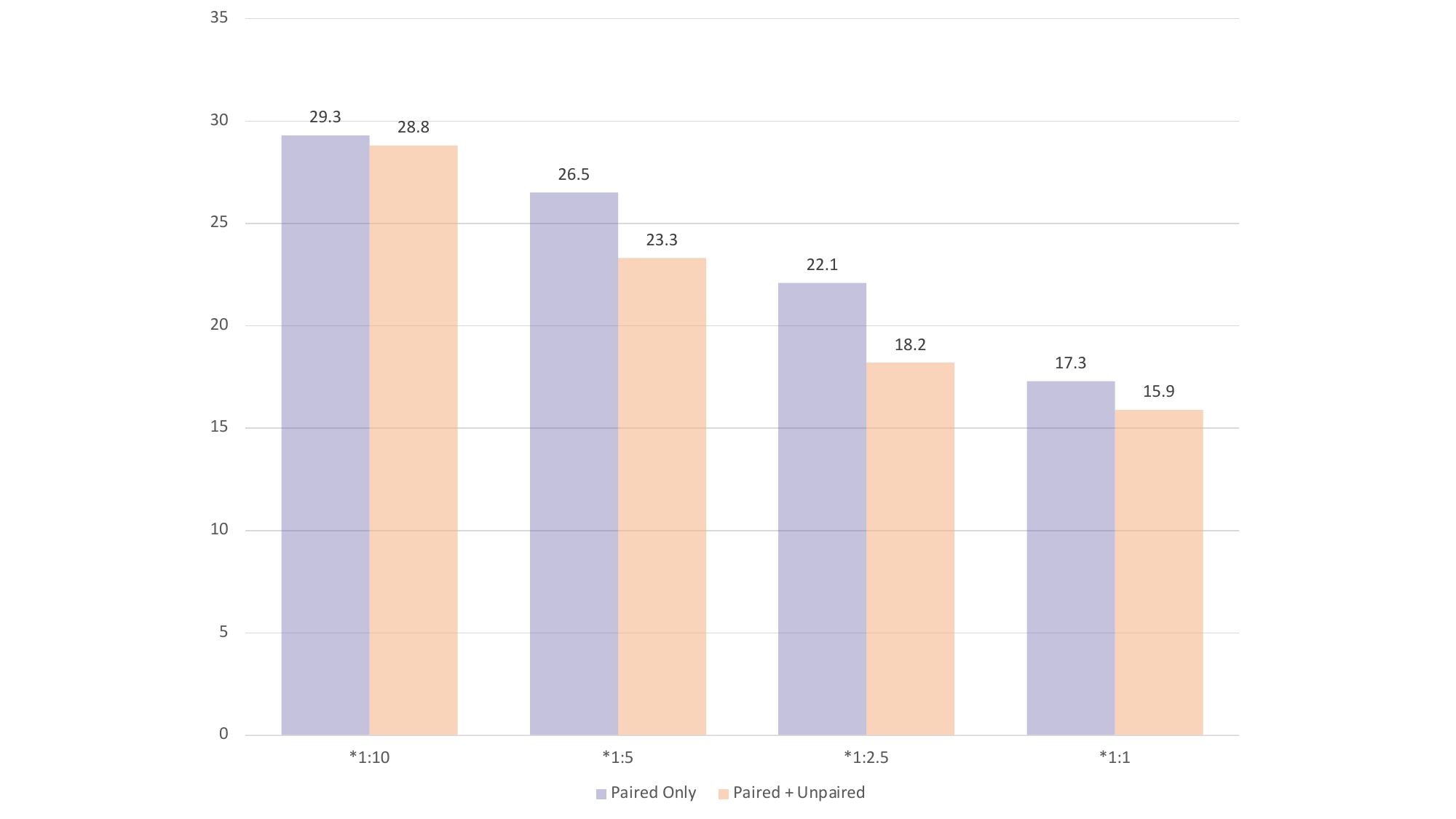} 
} 
\caption{Four groups of training data with different P/U ratios. The purple bars represent training with P data only, while the orange bars represent the results from training with P\&U data.} 
\label{fig:ratio} 
\end{figure}

We also compare the speech synthesis performance of the proposed model using only paired data and using paired data along with unpaired data, evaluate using PER. The results shown in Fig. \ref{fig:ratio} indicate that adding unpaired data to the training set reduces the PER, particularly in cases where there is insufficient paired data. Unpaired data has a more pronounced positive impact on our proposed model compared to the baseline model. This suggests that DQR-TTS can effectively utilize unpaired data to expand phoneme coverage of the dynamic codebook and address low-resource scenarios.

\subsection{Ablation Study}
In order to evaluate the influence of the semi-supervised training approach for text-to-speech synthesis, along with the dynamic codebook and its corresponding dynamic codebook update strategy, we carry out ablation experiments to evaluate the speech synthesis capability of DQR-TTS without these enhancements. In our ablation study experiments, we conduct experiments on the codebook and semi-supervised method, three cases are taken into consideration: 
\begin{enumerate}
\item Training the proposed model in a fully-supervised way.
\item Using a static codebook and training in a semi-supervised way, this means that the size of the codebook no longer increases during training with unpaired data.
\item Our proposed model, which contains a dynamic codebook and is trained in a semi-supervised way.
\end{enumerate}
Note that  case 1) is equal to \textit{w/} static codebook, fully-supervised. Training data for case 1) consists of 120 min of paired data, while for the latter two cases, training data comprises 120 min of paired data along with 300 min of unpaired data. For the models in case 2) which integrates a static codebook, we fixed the size of the codebook at the initial size of the dynamic codebook. Table \ref{tab:ablation} displays the outcomes of the ablation study.

\begin{table}[!htbp]
\centering
\caption{Ablation Study}
\label{tab:ablation}
\begin{tabular}{ccc}
\toprule
Case & MOS $\uparrow$ & MCD $\downarrow$\\
\midrule
Fully-supervised & 2.68 $\pm$ 0.24 & 6.11 $\pm$ 0.08 \\
{\itshape w/} static codebook, semi-supervised & 2.73 $\pm$ 0.10 & 5.93 $\pm$ 0.17 \\
\midrule
\textbf{DQR-TTS} & 3.06 $\pm$ 0.29 & 4.79 $\pm$ 0.06 \\
\bottomrule
\end{tabular}
\end{table}

For the ablation study, we can find that using a dynamic codebook improves the fidelity of synthesized speech. When using a static codebook, the performance of the model in case 2) does not show significant improvement compared to case 1) that with less training data. However, when using a dynamic codebook (\textit{i.e.} DQR-TTS), the caliber of generated speech improves significantly, indicating that the static codebook fails to make full use of unpaired data. This is likely because the fixed-size codebook in the experiment is insufficient to cover the phonemes present in the unpaired data adequately. The findings indicate that DQR-TTS is beneficial for fully leveraging unpaired data to improve synthesis quality of TTS under condition of limited resources.

\section{Conclusion}
\label{sec:typestyle}

In this paper, we introduce an innovative semi-supervised model for text-to-speech synthesis called DQR-TTS. The proposed model can cope with low-resource situations. As the crucial part of the proposed model, dynamic quantized representation module is incorporated into a sequential autoencoder, and contains a dynamic codebook. Experiments show that in low-resource scenario, our proposed model trained with limited paired data outperforms previous works in both subjective and objective metrics.

\section{Acknowledgement}
Supported by the Key Research and Development Program of Guangdong Province (grant No. 2021B0101400003) and corresponding author is Xulong Zhang (zhangxulong@ieee.org).

\clearpage
\newpage

\bibliographystyle{IEEEtran.bst}
\bibliography{refs.bib}

\end{document}